# Analysis of delay correlation matrices

K. B. K. Mayya and R. E. Amritkar

*Abstract*—We construct and analyze symmetrized delay correlation matrices for empirical data sets for atmopheric and financial data to derive information about correlation between different entities of the time series over time. The information about correlations is obtained by comparing the results for the eigenvalue distribution with the analytical results for the independent, identically distributed random data sets. For the atmospheric case we find long term correlations between different entities of the multivariable time series. For the financial time series we find little correlations between different entities over a time delay beyond about two days. Most of the eigenvalues for the symmetrized delay correlation matrices for the financial data are symmetrically distributed about zero. The delay correlation results for the financial data are similar to the analytical results for the random data sets. However there are considerable deviations for the atmospheric data from the random case.

## I. Introduction

THE study of multivariable time series is of importance in many fields of research. The methodology to study a multivariable time series [1] generally involves the construction of appropriate correlation matrices and their analysis by using different tools like singular value decompostion, factor analysis etc.

Recently random matrix methods are being increasingly applied to the study of multivariable data. Random matrix theory was introduced by Wignar, Dyson and Mehta to understand the complex systems [2], [3] such as atomic nuclie, many-body quantum systems etc. These methods were devised to obtain insight into the spectrum and transition strengths of complicated systems.

Recently these methods have been used to study the behaviour of multivariable systems like atmospheric variables and stock marktet data. The empirical correlation matrices constructed for the basic atmosphereic parameters have been found to satify random matrix prescriptions to a remarkable extent [4]. The physically relevant information may be derived out in this case by observeing the eigenmodes of the largest eigenvalues which deviate significantly from the random matrix prediction.

In the case of financial data, a major portion of the eigenvalue spectrum for the correlation matrix of price fluctuations shows a remarkable agreement with the theoretical prediction based on the random matrix theory [5]. This portion of the eigenvalue spectrum is found to satisfy the universal properties of the Gaussian orthogonal ensemble [7]. However it has been observed that there are serious discrepencies in the use of empirical correlation matrices for risk management due to the remaining portion of the spectrum which corresponds to the correlations between different variables [6].

In all the above cases the empirical correlation matrices are constructed and their properties are compared with the universal properties of certain random matrix ensembles satisfying certain symmetry properties. The empirical correlation matrices evaluate the equal time correlations between different entities in the given multivariable time series. These give us valuable information regarding the correlations between various entities of the system under consideration.

The above analysis using random matrix results however give little information about correlations between different entities of the multivariable time series at different times. A particular entity being measured by one of the variables of the time series may have some effect on another at a later time. This effect is not amenable to measurement or quantification using the empirical correlation matrices. This is because the corrlations between different entities in the correlation matrices are equal time correlations. To understand and quantify the effect of one variable or entity on another at a later time, one method would be to calculate the delay correlations i.e., correlations between one variable in the multivariable time series and another at a time delay $\tau$. We can then construct a matrix of such delay correlations and analyze their properties.

In our work we concentrate on such delay correlation matrices. We build delay correlation matrices for two different empirical data sets. First is the atmospheric data for monthly mean sea level pressure for the north atlantic region and the other is for the S&P 500 stock market data for the years 1991 to 1996. For these data sets we try to derive information on correlations over time between different entities of the multivariable time series. We also derive analytical results for independent, identically distributed random data sets. The information about correlations with time in the empirical data sets is obtained by comparing the empirical results with the analytical results for the random data sets.

## II. Construction of symmetrized delay correlation matrices

As mentioned above, the previous methods involved equal time correlations. The construction of delay correlation matrix involves calculating correlations between different entities with a time delay. Consider the multivariable time series at hand represented as a matrix **M** of order $N$x$T$. Here $N$ is the number of time series of length $T$ each. Suppose $i$ and $j$ are two time series among the given multivariable time series **M**. The correlation between $i$ at say $t = 0$ and $j$ at a time lag $t = \tau$ is given by

$$C_{ij} = \frac{1}{T} \sum_t M_{it} M_{jt+\tau} \qquad (1)$$

The matrix **C** thus constructed is asymmetric. The eigenvalues of such a matrix will be complex. To have real eigenvalues we suitably symmetrize the matrix **C**. The symmetrized matrix **C$^s$** is constructed according to the expression,

$$C^s_{ij} = C^s_{ji} = \frac{C_{ij} + C_{ji}}{2} \qquad (2)$$

K. B. K. Mayya (corresponding author) and R. E. Amritkar are in Physical Research Laboratory, e-mail: kbkmayya@prl.res.in.

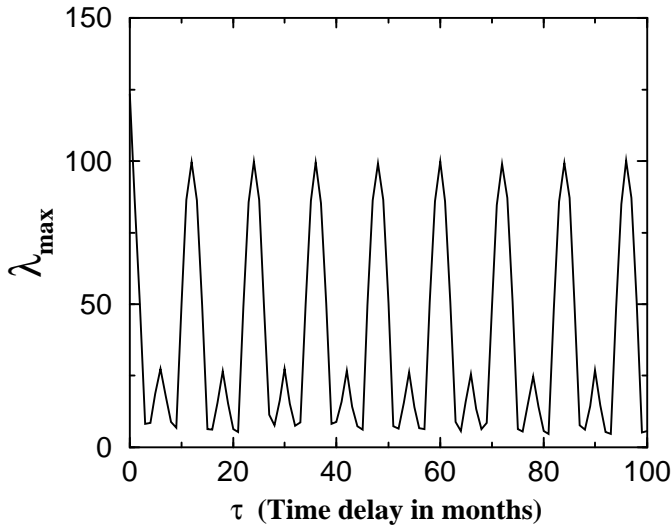

Fig. 1. Plot of $\lambda_{max}$ versus $\tau$ for the atmosphereic data. Here the ratio $Q = 1.435$

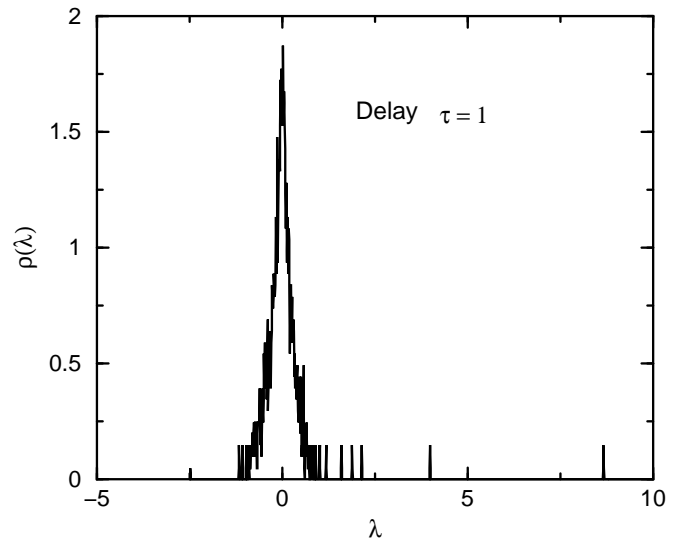

Fig. 2. Plot of $\rho(\lambda)$ versus $\lambda$ for $\tau = 1$ for the S&P 500 stock market data. Here the ratio $Q = 3.22$

The matrix element $M_{it}$ corresponds to the $t$th element of the time series $i$. The symmetrized delay correlation matrix $\mathbf{C^s}$ may be thus represented in terms of the matrix $\mathbf{M}$ by the expression

$$\mathbf{C^s} = \frac{\mathbf{M}^\dagger(0)\mathbf{M}(\tau) + \mathbf{M}^\dagger(\tau)\mathbf{M}(0)}{2T} \quad (3)$$

### III. NUMERICAL STUDIES

In the case of the atmospheric data, we use the monthly averaged mean sea level pressure data for the period 1948 to 1999. This data is collected for the North Atlantic domain i. e., $0^o - 90^o$ N, $120^o$ W - $30^o$ E. These data were obtained from the online database of National Centers for Environmental Prediction (NCEP) reanalysis archives. Each data is of length $T = 624$. The averaged mean sea level pressure is collected for $N = 434$ grid points. We define the ratio $Q = (T - \tau)/N$ (to be used later) which for the present case is 1.435.

We construct symmetrized delay correlation matrices for this data for different delay values $\tau$. Here $\tau = 1$ corresponds to one month. We take delay values with range $\tau = 1, ...100$. The symmetrized delay correlation matrices are constructed for each of the delay values and a plot of $\rho(\lambda)$ versus $\lambda$ is made for each of them. It is found that the eigenvalues are distributed about $\lambda = 0$. The eigenvalues oscillate about $\lambda = 0$ with the delay value $\tau$. To this end we make a plot of the largest eigenvalue $\lambda_{max}$ versus the time delay $\tau$. Figure (1) shows this plot of $\lambda_{max}$ versus time delay $\tau$. As we can see from the figure the $\lambda_{max}$ value shows oscillatory behaviour with a period of about one year.

Comparing with the results of random data sets (see section IV) we conclude that there is long term correlation among the various entities of the atmospheric data under consideration.

To this end we also analyze the cumulative distribution of the eigenvector components of the three largest eigenvlues. The plot of the cumulative distribution of eigenvector components (not shown) also reflect the periodic nature of the correlations over time of the multivariable time series for the mean sea level pressure data under consideration.

For the case of financial data we take the stock market data for the S&P 500 companies. We take daily data for 406 companies of the S&P 500 for the period 1991-1996. Each company data is of length $T = 1309$ days. The value of $Q$ here is 3.22.

We build a symmetrized delay correlation matrix for daily price fluctuations given by the expression

$$C_{ij}^s = \frac{1}{2T} \sum_{t=1}^{T} (\delta x_i(t)\delta x_j(t+\tau) + \delta x_i(t+\tau)\delta x_j(t)) \quad (4)$$

The $\delta x$'s are rescaled to have constant unit variance i.e.,

$$\sigma^2 = <\delta x_i^2> = 1 \quad (5)$$

The symmetrized correlation matrix is then diagonlaized to obtain the corresponding eigenvalues. The distribution of eigenvalues is then calculated. This above calculation is done for the delays $\tau = 1, ..., 100$. Figure (2) shows the plot of $\rho(\lambda)$ versus $\lambda$ for the delay value $\tau = 1$. As can be seen from the figure, most of the eignvalues fall within the range $\lambda = -0.925$ to $\lambda = 0.925$ and they are symmetrically distributed about $\lambda = 0$. This is in agreement with the analytical results for the random data sets (see section IV).

Very few eigenvalues fall beyond this range. The number of eigenvalues lying beyond this range becomes very small for delay $\tau > 2$. Figure (3) shows the plot of $\lambda_{max}$ versus $\tau$ for the stock market data. As can be seen the largest eigenvalue has a very low value for the matrices constructed for delay values $\tau$ greater than about two days. Therefore it can be concluded that there is little correlation among different entities of the multivariable time series for the stock market data beyond two days.

As a further analysis we study the eigenvector components of the largest eigenvalue. The distribution of eigenvector components of the largest eigenvalues (not shown) agree with the theoretically predicted Porter-Thomas distribution for random

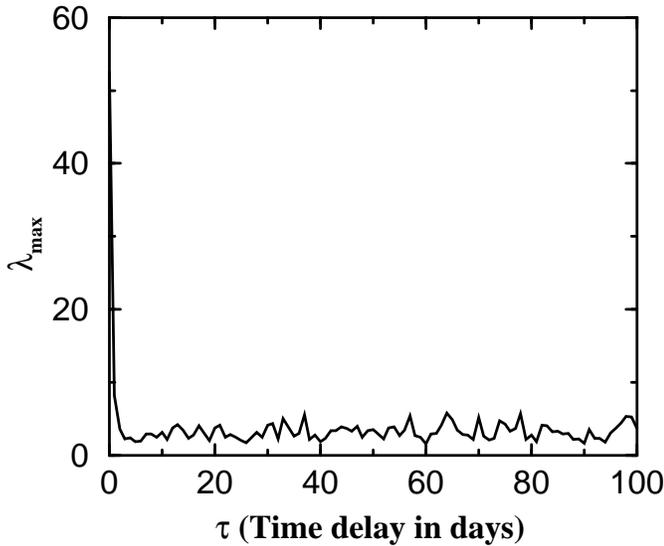

Fig. 3. Plot of $\lambda_{max}$ versus $\tau$ for the S&P 500 stock market data.

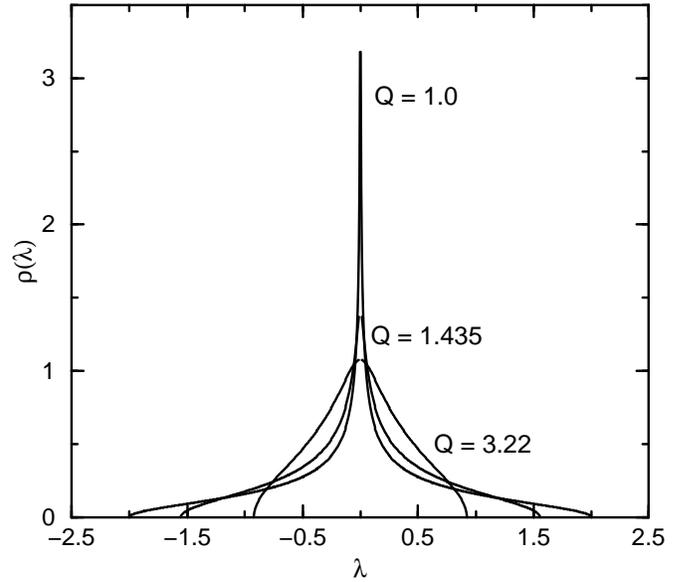

Fig. 4. Plots of $\rho(\lambda)$ versus $\lambda$ for thereee different values of $Q = 1$, $Q = 1.435$ and $Q = 3.22$.

matrices in the case of the correlation matrices constructed with delays greater than two days.

## IV. ANALYTICAL WORK

From the numerical results we observe that for the case of indenpendent, identically distributed data matrix **M**, the distribution of eigenvalues for the symmetrized delay correlation matrix is symmetric about $\lambda = 0$. To the best of our knowledge the behaviour of eigenvalues of the symmetrized delay correlation matrices have not been analyzed previously in the literature. To investigate this aspect further we carry out the following analytical work. The analytical calculation involves the derivation of the expression for the eigenvalue distribution for the given delay correlation matrix. The derivation for the normal correlation matrix is carried out in terms of the resolvent

$$G(z) = Tr[(z - \mathbf{M}^\dagger \mathbf{M})^{-1}] \quad (6)$$

where $G(z)$ is a complex function. The density of eigenvalues [5] is given by

$$\rho(\lambda) = \sum_n \delta(\lambda - \lambda_n) = \frac{1}{\pi} \lim_{\epsilon \to 0} Im[G(\lambda - i\epsilon)] \quad (7)$$

The above expression is valid for a simple correlation matrix for independent, identically distributed random variables. For the symmetrized delay correlation matrix $\mathbf{C^s}$, the derivation is done by using the resolvant in the form

$$G_\tau(z) = Tr[(z - C^s(\tau))^{-1}] \quad (8)$$

The expression for $G_\tau(z)$ is obtained by expanding the resolvant in powers of $1/z$ and using diagrammatic technique to represent various terms in the expansion. In the limit $T, N \to \infty$, while maintaining the ratio $Q = (T-\tau)/N$ to be a constant only the planar diagrams contribute [5]. We can sum the diagrams to infinite order and for $\tau \ll N$, we get the following fourth order equation for $G_\tau(z)$

$$G^4 + 2\kappa G^3 + (\kappa^2 - \frac{Q^2}{\sigma^4})G^2 - 2\kappa\frac{Q^2}{\sigma^4}G + \frac{Q^2}{\sigma^4\lambda^2}(2Q-1) = 0 \quad (9)$$

In the above equation $G_\tau(z)$ is represented as $G$ for convenience and

$$\kappa = \frac{Q-1}{\lambda} \quad (10)$$

Equation (9) is solved numerically to get the required solution for $G$. The imaginary part of the solution for $G$ is substituted in eqn. (7) to get the eigenvalue distribution for the delay correlation matrix.

Figure (4) shows the plot of $\rho(\lambda)$ versus $\lambda$ for three different values of $Q$ ($Q = 1.0$, $Q = 1.435$ and $Q = 3.22$). We see that $\rho(\lambda)$ is symmetrically distributed about $\lambda = 0$ for all values of $Q$.

The analytically obtained eigenvalue distributions are obtained in the limit $T, N \to \infty$. For the finite value of $T$ and $N$, we construct delay correlation matrices **M** for independent, identically distributed (Gaussian) random variables. We construct the symmetrized delay correlation matrix $\mathbf{C^s}$ for this matrix **M** of Gaussian distributed random variables and diagonalize it to get the eigenvalues. We calculate the distribution of eigenvalues of this symmetrized delay correlation matrix $\mathbf{C^s}$ and compare it with the distribution of eigenvalues obtained analytically for the appropriate $Q$ values. Figure (5) shows the plot of $\rho(\lambda)$ versus $\lambda$ for the numerically obtained distribution of eigenvalues for $Q = 1$. For comparison we plot the eigenvalue distribution obtained analytically for the case $Q = 1$ along with the numerically obtained distribution. As can be seen from the figure there is good agreement between the distribution of eigenvalues for the analytically and numerically obtained distributions.

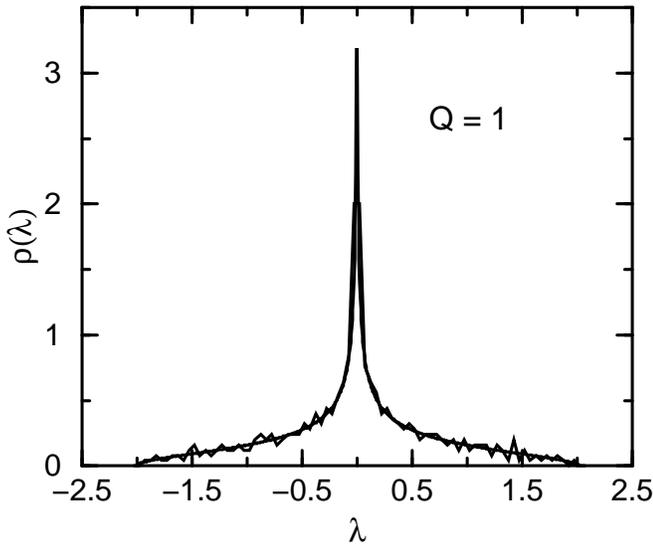

Fig. 5. Combined plot of $\rho(\lambda)$ versus $\lambda$ for $Q = 1$ obtained analytically as well as numerically for independent, identically distributed random data sets.

## V. Conclusions

To conclude we construct symmetrized delay correlation matrices for empirical data sets of atmospheric data and stock market data. We construct such matrices for varying delay values. The plot of largest eigenvalue versus the delay shows a periodicity of one year indicating that there is long term correlations over time among the different entities of the multivariable time series for the atmospheric data. For the financial data set we infer that there is little correlation between different stocks over time delays beyond about two days. For the financial data it is found that most of the eigenvalues are distributed symmetrically about zero. We have also obtained analytical distribution for the delay correlation matrices for the independent, identically distributed random variables. The eigenvalue distribution for the financial data shows similarity with the analytical distribution for random data sets, but the one for the atmospheric data shows considerable deviation. We can thus conclude that time correlations in financial data sets decay rapidly over a period of one or two days. After this period the data can be treated more or less as random. On the other hand the atmospheric data retains correlations for a very long time. The twelve month period is seen even after a delay of several years. The method proposed here can be used to analyze the correlations over time between the different variables of multivariable time series. Such multivariable time series are observed in several areas such as astrophysical data, biological data, EEG etc.

We thank Leloux et. al. [6] for providing us with the the S&P 500 stock market data. Also we thank Dr. M. S. Santanam [4] for providing us the atmospheric sea level pressure data downloaded from the North Atlantic region downloaded from http://www.cdc.noaa.gov maintained by NOAA-CIRES Climate Diagnostics Center, Boulder, CO.